\documentclass[showpacs,amsmath,amssymb,aps,10pt,reprint,superscriptaddress]{revtex4-1}
\usepackage{graphicx}
\usepackage{bm}
\usepackage[breaklinks=true,colorlinks=true,linkcolor=blue,urlcolor=blue,citecolor=blue]{hyperref}
\usepackage{graphics}
\usepackage{dcolumn}
\usepackage{amsmath,amssymb}
\usepackage{mathdots}
\usepackage{natbib}
\usepackage{soul,color}
\usepackage{epstopdf}
\usepackage[note-name]{notes2bib}

\begin{document}
\title{Role of magnons and the size effect in heat transport through an insulating ferromagnet/insulator interface}

\author{Valerij~A.~Shklovskij}
\affiliation{Physics Department, V. Karazin Kharkiv National University, 61077 Kharkiv, Ukraine}
\author{Volodymyr~V.~Kruglyak}
\affiliation{School of Physics and Astronomy, University of Exeter, EX4 4QL Exeter, UK}
\author{Ruslan V. Vovk}
\affiliation{Physics Department, V. Karazin Kharkiv National University, 61077 Kharkiv, Ukraine}
\author{Oleksandr~V.~Dobrovolskiy}
\email[Corresponding author: ]{Dobrovolskiy@Physik.uni-frankfurt.de}
\affiliation{Physikalisches Institut, Goethe University, 60438 Frankfurt am Main, Germany}
\affiliation{Physics Department, V. Karazin Kharkiv National University, 61077 Kharkiv, Ukraine}

\date{\today}

\begin{abstract}
While recent experiments on the spin Seebeck effect have revealed the decisive role of the magnon contribution to the heat current $Q$ in hybrid systems containing thin ferromagnetic layers, the available acoustic mismatch theory does not account for their magnetic properties. Here, we analyze theoretically the heat transfer through an insulating ferromagnet (F) sandwiched between two insulators (I). Depending on the relation between the F thickness, $d$, and the mean free path of phonons generated by magnons, $l_{ls}$, we reveal two qualitatively different regimes in the nonlinear heat transport through the F/I interfaces. Namely, in thick F layers the regime of conventional ``Joule'' heating with $Q \propto T_s^4$ is realized, in which the detailed structure of the F/I interfaces is inessential. Here $T_s$ is the magnon temperature. By contrast, in thin F layers with $d\ll l_{ls}$, most of phonons emitted by magnons can leave F without being absorbed in its interior, giving rise to the \emph{magnon overheating} regime with $Q \propto T_s^m$ and $m\gtrsim7$. Conditions for the examination of both regimes and the determination of $T_s$ from experiments are discussed. The reported results are relevant for the theoretical analysis of the spin Seebeck effect and the development of magnon-based spin caloritronic devices.
\end{abstract}
\pacs{65.40.-b, 75.30.Ds, 63.20.kd, 63.20.kk}
\maketitle

\section{Introduction}

When heat passes through an interface between two dissimilar solids, scattering of heat carriers at the boundary between them leads to a temperature jump $\Delta T = T_1 -T_2$, where $T_1$ and $T_2$ are the temperatures of the substances. This temperature jump appears in consequence of the thermal boundary resistance, discovered by Kapitza at boundaries of solids emersed in liquid helium \cite{Kap41jph,Pol69rmp}, and known as \emph{Kapitza resistance}. Within the framework of the acoustic mismatch theory, Little showed \cite{Lit59cjp} that at low temperatures $T\ll\Theta_D$ the heat current through the interface between two media is expressed by $Q=A({T_1}^4 -{T_2}^4)$. Here, ${\Theta_D}$ is the Debye temperature and the coefficient $A$ is determined by the acoustic properties of the contacting substances. If phonons hit the interface at oblique angles $\theta$, then $A$ is proportional to the interface transparency $\alpha(\theta)$ averaged over the angles $\theta$, a quantity representing the probability that a given phonon will pass through the interface between the two media. In the \emph{linear} approximation in $T$, from Little's result follows the Newton relation $Q =(R_{th}^{-1})\Delta T$, where $R_{th} (T) \approx (4AT^3)^{-1}$ is the Kapitza resistance $R_{th}$ which increases as $\sim 1/T^3$ with decreasing temperature.

The last decade has been marked by a growing interest in the generation of pure spin currents in spintronics \cite{Wol01sci,Zut04rmp,Sin12nam} and spin caloritronics \cite{Bau12nam,Sch13prb,Boo14ees}. The latter domain combines thermoelectrics with spintronics and nanomagnetism and it is concerned with the interplay between spin and heat currents in materials. The spin current may be formed by charge currents with opposite flow directions for spin up and spin down carriers, or it can consist of magnons, the quanta of collective spin excitations \cite{Kaj10nat}. In particular, the spin current is an inherent ingredient in spin pumping \cite{Tse02prl,Mos10prb}, the longitudinal spin Seebeck effect (LSSE) \cite{Uch08nat,Uch10nam,Pra18prb} where the spin current flows along the thermal gradient in the magnetic material, and the spin Hall magnetoresistance \cite{Nak13prl,Alt13prb,Vli13prb}. These effects have been extensively studied experimentally \cite{Uch08nat,Uch10nam,Nak13prl,Pra18prb} and theoretically \cite{Tse02prl,Xia10prb,Jia13prl,Tik13nac}, both taken alone as well as in comparison \cite{Wei13prl,Agr14apl,Agr14prb}. From a theoretical point of view, all these effects are governed by the generation of a current of angular momentum via a nonequilibrium process. Furthermore, in the field of magnon spintronic \cite{Chu15nph}, concerned with structures, devices and circuits that use spin currents carried by magnons, the quanta of spin waves (magnons) are used to carry and process information as alternative to charge-current-driven spintronic devices \cite{Chu14nac,Gru16nan}. Recently, pure magnonic spin currents in insulating ferromagnets have been suggested for the implementation of efficient logic devices \cite{Cra18nac}. At the same time, spin waves can transport heat in the same manner as the lattice excitations (phonons) transport heat through perturbations of the atom positions \cite{San77prb,Ant13nam}.

In the context of recent research, our study of the nonlinear heat transport across an F/I interface has been motivated by two experimental works on the longitudinal SSE \cite{Sch13prb,Guo16prx}. Namely, the authors of Ref. \cite{Sch13prb} calculated the phonon, electron, and magnon temperature profiles in YIG/Pt bilayers \bibnote{See Fig.~3 in \cite{Sch13prb} where the phonon, electron and magnon temperature profiles are shown for an insulator/insulating-ferromagnet/normal-metal structure} taking into account the thermal boundary resistances in the linear approximation. The analysis \cite{Sch13prb} has revealed that in thin-film hybrids the magnetic heat conductance qualitatively affects the magnon temperature and especially for magnetic insulators the determination of the phonon temperature profile is of central importance. The other work \cite{Guo16prx} was devoted to a study of the temperature-dependent SSE in heavy-metal/YIG hybrid structures as a function of the YIG thickness and the magnetic field strength for different heavy metal layers. The SSE signal exhibited a pronounced peak at low temperatures, and the SSE peak temperature strongly depended on the film thickness and the magnetic field. These results can be explained well within the framework of the magnon-driven SSE by taking into account the temperature-dependent effective propagation length of thermally excited magnons, which is also discussed in recent work \cite{Pra18prb}. In this way, the experimental results \cite{Sch13prb,Guo16prx,Pra18prb} emphasize the decisive role of interface effects in the temperature-dependent SSE and that magnon energy relaxation mechanisms by the phononic environment must be invoked generally for a complete understanding of thermal spin transport, and particularly for the physics underlying the SSE. The magnon free path is crucial for the understanding of the general peculiarities of the magnon-phonon interaction \cite{Uch11nam,San11prl,Ruc14prb,Boo14prb,Kik16prl,Boz17prl,Noa18jpd} as for the engineering of SSE spin-caloritronic devices \cite{Jun13apl,Lan17prb}.

As the same time, for the interpretation of experiments on the heat transport through an F/I boundary the acoustic-mismatch theory \cite{Lit59cjp} is usually applied, in which the temperature discontinuity at the interface is determined only by the acoustic characteristics of the media. Here, we show that the acoustic-mismatch theory is valid only for sufficiently thick insulating ferromagnets. Of primary interest to us is the opposite limiting case of F thin films where the role of magnons in the formation of the temperature discontinuity at the F/I interface becomes decisive. In particular, we show that at $T\ll\Theta_D$ a \emph{size effect} exists for the \emph{nonlinear} heat current $Q$ crossing an insulating ferromagnet/insulator (F/I) interface. Namely, $Q$ depends on the magnetic properties of \emph{thin} heated F films and the role of magnons is essential in the heat transfer regime which we term \emph{magnon overheating}. By contrast, for \emph{thick} heated F layers $Q$ can be described by the Little formula which does not account for the magnetic properties of the F layer.

\section{Problem statement}
\begin{figure}[t]
    \centering
    \includegraphics[width=1\linewidth]{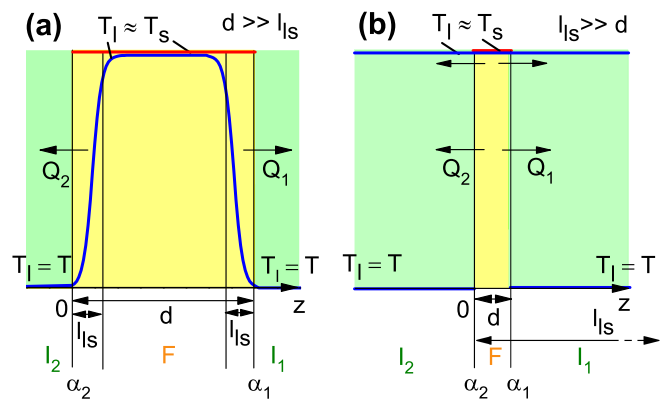}
    \caption{Size effect in the considered problem: An insulating ferromagnet is sandwiched between two insulators $I_1$ and $I_2$. Two regimes are considered: (a) The mean free pass $l_{ls}$ of thermal phonons generated by magnons is much smaller than the thickness of F. Only phonons within thin layers of the order of $l_{ls}$ at the interfaces may leave F and get absorbed in I$_i$. This regime corresponds to the ``Joule'' heating, described well by the Little approach \cite{Lit59cjp}. (b) The thickness $d$ of F is much smaller than $l_{ls}$. Most of phonons emitted by magnons leave F without being absorbed in its interior and do not return into F. This case corresponds to the magnon overheating regime which of primary interest to us. In both cases, the magnons existing within F are characterized by the magnon temperature $T_s$. The phonon temperature in $I_1$ and $I_2$ is $T_l = T$. The Kapitza jump at the boundaries of the thick F layer in (a) does not depend on the properties of F while it depends on the properties of the thin F layer in (b). The interfaces are characterized by the transparency coefficients $\alpha_i(\theta)$, where $\theta$ is the angle between the phonon wave vector and the $z$ axis.}
    \label{f1}
\end{figure}

We consider an insulating ferromagnetic film F of thickness $d$ sandwiched between two bulky insulators I$_1$ and I$_2$, whose temperatures are known Fig. \ref{f1}. We choose the $z$ axis perpendicular to the interfaces of the media and assume the problem to be spatially homogeneous in the $oxy$ plane. The magnon temperature $T_s$ in F is constant over the film thickness, i.e. along the $z$ axis, as will be justified next. Specifically, we consider the stationary case which can be realized in two different ways. The first case is when the temperatures of I$_1$ and I$_2$ are different, $T_1\neq T_2$. In the second case, illustrated in Fig. \ref{f1}, $T_1 = T_2 = T$ and the stationary heat current is supported by a parametric pumping with power $W$. The transparencies $\alpha_1$ and $\alpha_2$ of the F/I interfaces for the phonons are known. The task is to derive the heat currents through the interfaces $Q_1$ and $Q_2$.

The calculation of the heat currents in the system at arbitrary values of $\alpha_1$, $\alpha_2$ and $d/l_{ls}(T_s)$ is possible because of two physical circumstances simplifying the considered problem. Firstly, a homogenous magnon distribution over the film thickness can be assumed when ${K_s}/{K_p}\gg1$, where $K_s$ and $K_p$ are the heat conduction coefficients for the magnons and the phonons in F, respectively. In particular, this condition is justified at low temperatures when either $\Theta_C\gg\Theta_D$ leads to $K_s/K_p \sim \sqrt{\Theta_D \Theta_C^4/T^5}\gg 1$ for $T_s\ll{\Theta_D}^2/\Theta_C$  \cite{Akh67boo} or $K_s/K_p \sim \sqrt{ \Theta_D^6/T^3 \Theta_C^3}\gg1$ for $\Theta_D\gg\Theta_C$ \cite{Akh58etp}. Here, $\Theta_C$ is the Curie temperature. Secondly, even when the magnon temperature of the Boze-Einstein distribution can no longer be established on the basis of direct intermagnon collisions, $T_s$ can still be introduced  \cite{Akh67boo}. Namely, the value of $T_s$ is justified in the limit $d\gg l_{ls}$ because of the effective intermagnon collisions via the phonons. These two circumstances allow us to reduce the formulated problem to a solution of the stationary kinetic equation for the phonon distribution function, and then to determine $T_s$ as a function of $Q$ and the temperatures of the insulators from the heat-balance equation.

In addition, the good-transparency case $\alpha \sim 1$ will be of especial interest to us as it allows for simple boundary conditions for the phonon distribution function. The ballistic propagation of the phonons emitted by the F film not only simplifies the expressions for the heat dissipation in the sample, but it also stipulates the necessary condition for the realization of the \emph{size effect}.

In accordance with the considerations above, we assume that the distribution of magnons is characterized by the temperature $T_s$. In the limit $d \ll l_{ls}$ the inhomogeneity of $T_s$ along the $z$ axis can be neglected because of ${K_p}/{K_s}\ll 1$. At the same time, the Bose-Einstein distribution for phonons $N_\mathbf{q}(z)$, where $\mathbf{q}$ is the phonon wave vector, can be essentially inhomogeneous and it should be determined from the kinetic equation
\begin{equation}
    \label{eKinEq}
    s_z \frac{\partial N_{\mathbf{q}}(z)}{\partial z} = L_{ls} \lbrace N, n\rbrace,
\end{equation}
with appropriate boundary conditions. In Eq. \eqref{eKinEq}, $s_z$ is the projection of the phonon velocity on the $z$ axis and $L_{ls}$ is the phonon-magnon collision integral \cite{Akh67boo}, which can be expressed as
\begin{equation}
    \label{eColInt}
     L_{ls} \lbrace N, n\rbrace = \nu_{ls}(T_s, q) \lbrace n(T_s) - N_\mathbf{q}(z)\rbrace.
\end{equation}
Here, $n(T_s)$ is the equilibrium Bose-Einstein distribution at the magnon temperature $T_s$ with the dispersion law $\varepsilon_k=\Theta_C{(ak)}^2$ in the long-wave $ka\ll 1$ limit, $a$ is the lattice constant, $k$ is the magnon wave number and $\nu_{ls}(T_s, q)$ is the average frequency of collisions between the phonons of frequency $\omega_q = sq$ and the magnons.

Adapting the solution scheme for the kinetic equation from Ref. \cite{Shk80spj}, with details placed in the Appendix section, we denote the phonon reflection coefficients at boundaries 1 and 2 as $\beta_1$ and $\beta_2$, such that $\beta_i = 1 - \alpha_i, i=1,2$, where $\alpha_i(\theta)$ is the transparency coefficient. We consider the case of ballistic propagation of the phonons emitted by F through the F/I boundary, taking into account the finite transparency of the F/I interface within the framework of the acoustic-mismatch theory \cite{Lit59cjp}. We introduce two new functions $N^\gtrless(\mathbf{q},z) = N(z,q,q_z\gtrless 0)$ allowing us to write the boundary conditions for $N(z)$ in Eq. \eqref{eKinEq} at $z = 0$ and $z = d$ as
\begin{eqnarray}
    \label{eBonCond}
    N^> (0) & = \alpha_1 n(T_1) + \beta_1 N^< (0), \nonumber\\
    N^< (d) & = \alpha_2 n(T_2) + \beta_2 N^> (d).
\end{eqnarray}
These boundary conditions are written for the assumed absence of reversed phonons in I$_1$ and I$_2$. This assumption is justified when I$_1$ and I$_2$ are bulk.

\section{Main results}
The solution of Eq. \eqref{eKinEq} allows us to express the heat current $Q=\sum_{\bf q}(\hbar\omega_{\bf q}) \dot{N}_{\bf q}$ crossing the respective interface as
\begin{equation}
    \label{eHeatFlux}
    \begin{array}{lll}
    Q_1 = \sum_{q_{z > 0}}[\alpha_1 \eta (1 - \beta_2  x^2) \lbrace n(T_1 ) - n(T_s )\rbrace \\[2mm]
    \hspace{40mm}-\alpha_2  x \lbrace n(T_2 ) - n(T_s )\rbrace],\\[3mm]
    Q_2 = -\sum_{q_{z < 0}}[\alpha_2 \eta (1 - \beta_1  x^2) \lbrace n(T_2 ) - n(T_s )\rbrace \\[2mm]
    \hspace{40mm}-\alpha_1  x \lbrace n(T_1 ) - n(T_s )\rbrace],\\
    \end{array}
\end{equation}
where $\eta=\hbar \omega_q s_z/(1 - \beta_1 \beta_2 x^2)$, $x = \exp(-d/l)$, $l \equiv |l_z|$ depends on the angle $\theta$ between the direction of the vector $\mathbf{q}$ and the $z$ axis. The value of $T_s$ is determined from the heat balance equation for the magnons $Q = Q_1 - Q_2$, where $Q=Wd$ is the total density of the heat currents passing through the interfaces.

The relation between $Q$, $T_s$ and $T_l$, which follows from \eqref{eHeatFlux}, can be essentially simplified in the case $T_1=T_2=T_l$
\begin{equation}
    \label{eHeatFluxComb}
    Q= \sum_{q_{z > 0}}\hbar \omega_q s_z \tilde\alpha(\mathbf{q},d) \lbrack n (T_s )- n (T_l )\rbrack,
\end{equation}
where $\tilde\alpha(\mathbf{q},d)$ is the combined interface transparency
\begin{equation}
    \label{eComTransp}
    \tilde\alpha \equiv \frac{(1-x)\lbrack \alpha_1 (1+\beta_2 x)+\alpha_2 (1+\beta_1 x)\rbrack}{(1-\beta_1 \beta_2 x^2)}.
\end{equation}
\begin{figure}
    \centering
    \includegraphics[width=0.8\linewidth]{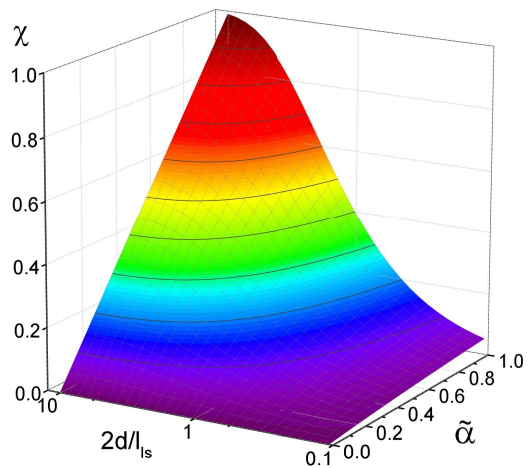}
    \caption{The form factor $\chi_Q(q)$ calculated by Eq. \eqref{eChi} as a function of the parameter $2d/l_{ls}$ and the combined transparency $\tilde{\alpha} = 2\alpha(1-x)/(1-\beta x)$ for the case $\beta_1 = \beta_2$.}
    \label{f2}
\end{figure}

We note that in contrast to the ``bare'' transparencies $\alpha_i(\theta)$, which depend only on the phonon incidence angle at the interface, $\tilde\alpha$ contains not only an additional angular dependence mediated by $x$, but it also depends on the phonon frequency. Proceeding in Eq. \eqref{eHeatFluxComb} from the sum to integration, the heat current $Q$ can be presented as
\begin{equation}
    \label{eHeatFluxInt}
    Q = \frac{2\pi s^2}{{(2\pi\hbar})^3} \int\limits_0^{q_D} {q^3}  dq \chi_Q(q)\lbrack n(\varepsilon_q/T_s ) - n(\varepsilon_q/T_l)\rbrack,
\end{equation}
where $\chi_Q(q)$ is the effective  transparency ``form factor'' averaged over the angles $\theta$ and defined as
\begin{equation}
    \label{eChi}
    \chi_Q(q) =\int\limits_0^1 {v}  \tilde\alpha(v,q) {dv}.
\end{equation}
Here $v = \cos\theta$, $q$ is the phonon wave number, $\varepsilon_q = sq$, and $\tilde\alpha(v,q)$ is given by Eq. \eqref{eComTransp}. In this way, Eqs. \eqref{eComTransp}--\eqref{eChi} link $Q(T_s)$ with the film thickness $d$ and the transparencies of the interfaces. In the general case, the dependence of the effective transparency ``form factor'' $\chi_Q(q)$ on the phonon momentum $q = {\hbar\omega_q}/s$ can be calculated by numerical integration. In the two limiting cases discussed in what follows, expressions for $\chi_Q(q)$ can be derived analytically.

To this end, we introduce the temperature-dependent parameter
\begin{equation}
    \label{eEpsilon}
    \varepsilon=\frac{2d}{l_{ls}}\frac{\beta_1 \beta_2}{(1-\beta_1 \beta_2)},
\end{equation}
where $l_{ls} = s/\nu_{ls}(T_s, q)$ and
\begin{equation}
    \label{eNuls}
    \nu_{ls}(T_s, q) = D(T_s)\sum_{p=1}^{\infty} (1-e^{-px}){\int_{y_0}^\infty}y (y+x) e^{-py} dy
\end{equation}
has a physical meaning of the inverse lifetime of a phonon with frequency $\omega_q$ with respect to the absorption or emission of the phonon by a magnon.

\begin{figure}[t]
    \centering
    \includegraphics[width=0.9\linewidth]{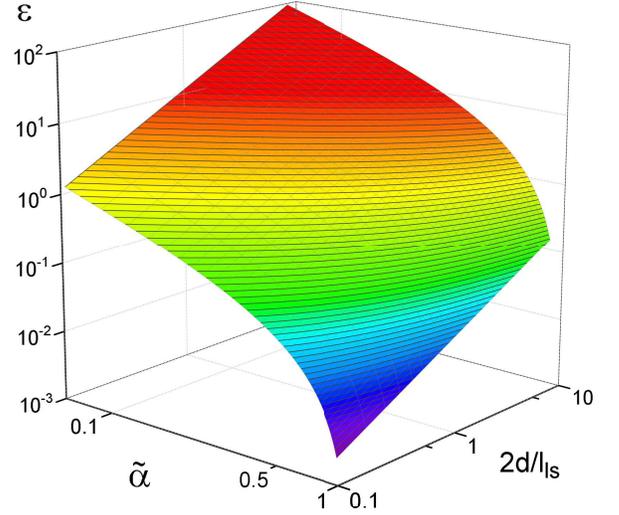}
    \caption{Dependence of the parameter $\varepsilon$ on the interface transparency and the ratio $2d/l_{ls}$ by Eq. \eqref{eEpsilon} when $\beta_1=\beta_2$. While the conventional regime of ``Joule'' heating is realized for $\varepsilon \gg 1$ in thick films with $d \gg 2d/l_{ls}$ and/or bad interface transparencies $\alpha\ll 1$, the magnon overheating regime ensues for $\varepsilon \ll 1$ in thin films with $d \ll 2d/l_{ls}$ and good interface transparencies $\alpha\approx1$.}
    \label{f3}
\end{figure}

The dependence of the parameter $\varepsilon$ on the interface transparency and the ratio $2d/l_{ls}$ is plotted in Fig. \ref{f3}. The ``bulk'' case, in which the magnon contribution to the thermal boundary resistance is neglected is obtained from Eq. \eqref{eHeatFluxInt} and \eqref{eChi} in the limiting case $\varepsilon\gg 1$. Indeed, in this limit $\tilde\alpha = \alpha_1 + \alpha_2 =\bar\alpha$ and at $T_s\ll\Theta_D$ the well-known Little result follows from Eq. \eqref{eHeatFluxInt}
\begin{equation}
    \label{eLittle}
    Q = Wd = A({T_s}^4- {T_l}^4),
\end{equation}
where the constant $A={\pi^2 \bar\alpha}/{s^2 \hbar^3}$ is determined only by the acoustic characteristics of F. We note that if $W$ is constant, then $T_s\sim d^{1/4}$, i.e. it weakly increases with increasing thickness of F. This thermal regime corresponds to the conventional \emph{``Joule'' heating}. In particular, the spectrum of phonons emitted by F is described by the equilibrium temperature $T_s$, so that the maximum of its spectral intensity corresponds to the energy $\hbar\omega_m \sim 2.8 T_s$, where $T_s$ is in energy units.

In the opposite limiting case $\varepsilon\ll1$ we consider thin films $d\ll l_{ls}$ with $d >\lambda$, where $\lambda = \lambda(T_s)$ is the wavelength of thermal phonons in F, when the deformation of the phonon spectrum in F can be neglected. In this case, from Eq. \eqref{eComTransp} we obtain $\tilde\alpha \approx 2d/l$ and $\chi_Q(q)=2d/l_{ls}(T_s)$. It can be shown that Eq. \eqref{eHeatFluxComb} is reduced to
\begin{equation}
    \label{eHeatFluxThinFilm}
    Q=Wd=d \sum_{q_{z > 0}}\hbar \omega_q \nu_{ls}(T_s, q) \lbrack n (T_s )- n (T_l )\rbrack.
\end{equation}

As it is seen from Eq. \eqref{eHeatFluxThinFilm}, in this case $T_s$ \emph{does not depend on the transparency of the interfaces at all}. Most of phonons radiated by magnons manage to leave the film as they are not reabsorbed inside the film. Accordingly, magnons and the lattice are described by two different temperatures $T_s$ and $T_l$, $T_s> T_l$, corresponding to the regime of \emph{magnon overheating}. The result of the calculation of the heat current by Eq. \eqref{eHeatFluxThinFilm} is
\begin{equation}
    \label{eW}
    W=\Phi K(p),
\end{equation}
where
\begin{equation}
    \label{ePhi}
    \Phi = \frac{N}{(2\pi)^3} \frac{\Theta_D^2}{2\hbar}
    \frac{\Theta_C}{\Theta_p}(T_s/\Theta_C)^3[(T_s/\Theta_D)^4 -(T_l/\Theta_D)^4]
\end{equation}
and $\Theta_p=Ms^2$. The function $K(p)$ is given \cite{Shk18arx} by
\begin{equation}
    \label{16}
    K(p)={\int_0}^\infty \frac{ u^3 du}{e^x -1} [J_D (T_s, x=u,y_0)- J_D(T_s, x=u/\gamma,y_0)],
\end{equation}
where $\gamma=T_s/T_l$ is the magnon overheating parameter
\begin{equation}
    \label{17}
    \begin{array}{lll}
    [J_D (u) - J_D(u/\gamma)]=\\[2mm]
    \sum_{p=1}^\infty u\phi_1[(1-e^{-pu})-(1/\gamma)(1- e^{-pu/\gamma}]\\[2mm]
    +\phi _2[(1-e^{-pu})-(1-e^{-pu/\gamma})].
    \end{array}
\end{equation}
Here, $\phi_1(p, y_0 )= e^{-py_0}(y_0/p +1/p^2)$, $\phi _2(p, y_0) = e^{-py_0} ({y_0}^2/p + 2y_0/p^2 + 2/p^3)$, and $y_0={{\Theta_D}^2}/{4T\Theta_C }$.

\begin{figure}[t]
    \centering
    \includegraphics[width=0.9\linewidth]{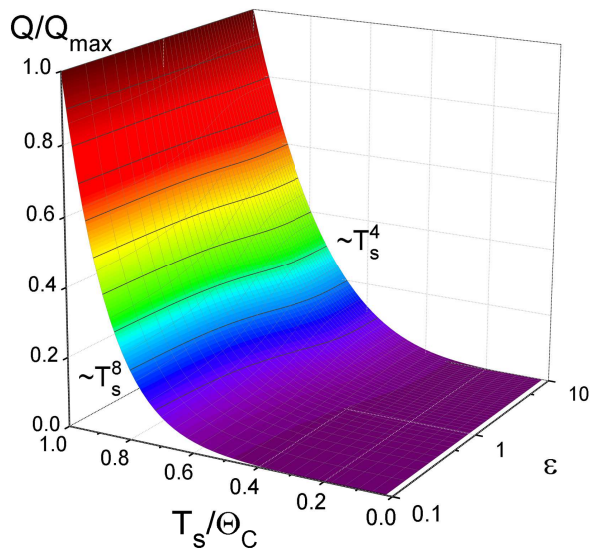}
    \caption{Size effect in the nonlinear heat current $Q$, normalized to its maximal value $Q_{max}$, given by Eq. \eqref{eHeatFluxInt} through the F/I interface: While in thick F layers with $\varepsilon\gg1$ the regime of ``Joule'' heating is described by the well-known Little result $Q \propto T_s^4$ \cite{Lit59cjp}, for thin F layers with a good interface transparency ($\varepsilon\ll1$) the dependence $Q \propto T_s^m$ with $m \gtrsim 7$ results in the magnon overheating regime. The plot is calculated for $\Theta_C = \Theta_D$.}
    \label{f4}
\end{figure}

The size effect in the nonlinear heat current $Q$ given by Eq. \eqref{eHeatFluxInt} through the F/I interface is illustrated in Fig. \ref{f4}. Namely, while in the limit of thick F layers with $\varepsilon\gg1$ the regime of ``Joule'' heating is described by the well-known Little result \cite{Lit59cjp} corresponding to $Q \propto T_s^4$ in Eq. \eqref{eLittle}, for thin F layers with a good interface transparency $\varepsilon\ll1$ the dependence $Q \propto T_s^m$ with $m \gtrsim 7$ results in the magnon overheating regime described by Eq. \eqref{ePhi}. While the exponent $m = 7$ for $T_s$ enters Eq. \eqref{ePhi}, an additional temperature dependence ($\propto T_s^n$, $n \lesssim 1$) is brought about by the function $K(p)$ \cite{Shk18arx}.

\section{Discussion}
\label{Discussion}

We proceed to a general discussion of the obtained results and begin with the elucidation of the physical origin of the increasing magnon contribution with decrease of the thickness of F. First, we consider qualitatively the mechanism of the heat transfer through the F/I interfaces in the I$_1$/F/I$_2$ trilayers sketched in Fig. \ref{f1}. Obviously, although magnons are the principal heat carriers in F, they cannot enter the I layers. This is why the heat transfer through the F/I boundaries is mediated by phonons and it depends on the acoustic transparency of the interfaces. In the F layer, near its boundaries, transition layers exist in which the thermal energy transported by magnons is transformed into the phonon flux. The thickness of these layers is on the order of the mean free pass $l_{ls}$ of the thermal phonons relative to their scattering on magnons. Then, if the thickness of the insulating F layer $d$ is much larger than the phonon mean free pass, $d\gg l_{ls}$, the detailed structure of the transitions layer becomes inessential for the calculation of the heat current through the F/I boundary. This corresponds to the conventional approach used by Little within the acoustic-mismatch theory \cite{Lit59cjp} in which the magnon contribution to the heat current is neglected. The size effect becomes apparent in the opposite case $d\ll l_{ls}$, when in the ballistic regime $\alpha \sim 1$ most of phonons emitted by magnons leave F without being absorbed in its interior even after several successive reflections from its boundaries. Therefore, in contrast to the previous case, the spectral distribution of the phonons emitted by the F film contains more information on the magnon-phonon interaction in the insulating F than on the transparency $\alpha$ of the F/I boundary.

To augment the appearance of the size effect, we analyze in more detail the expression for the phonon distribution function. For simplicity, we consider the symmetric case when $\alpha = \alpha_1 = \alpha_2$ and $T_l = T_1 = T_2$. Then, the expressions for $N^\gtrless$ acquire the form
\begin{equation}
    \label{eN><sym}
    \begin{array}{lll}
    N^>(z) = \varkappa  e^{- z/l}n(T) + (1 - \varkappa  e^{- z/l})n(T_s),\\[2mm]
    N^<(z) = \varkappa  e^{\frac{z - d}l}n(T) + (1 - \varkappa   e^{\frac{z - d}l})n(T_s),
    \end{array}
\end{equation}
where $\varkappa = \alpha/(1-\beta x)$. From Eqs. \eqref{eN><sym} it is seen that the characteristic length of the spatial variation of $N^\gtrless$ is $l ={|s_z|}/{\nu_{ls}} = l(q,\theta)$. The functions $N^\gtrless$ are ``weighted'' sums of two equilibrium Bose distributions. Namely, $n(T_l)$ stands for the ``cold'' phonons from the bath while $n(T_s)$ for the ``hot'' phonons emitted by magnons of F. The relative weight of these terms is determined by the quantity $\varkappa$. The physical meaning of $\varkappa$ is the effective probability, $0<\varkappa<1$, that a phonon, which is incident from F on the F/I interface, will leave F without collisions with magnons. It can be shown that even in the case of small bare transparencies $\bar\alpha\ll1$, the quantity $\varkappa$ can nevertheless be of the order of unity if $\xi\equiv d/l\ll\alpha$. This means that if $\xi\ll\alpha$, then practically all phonons emitted by magnons and satisfying this condition will leave F without being reabsorbed in its interior with a probability on the order of unity, despite of the possibility for a series of successive reflections from the film boundaries. In accordance with this scenario, from the definition of $\varkappa$ follows ${d\varkappa}/{d\alpha} > 0$ and ${d\varkappa}/{d\xi} < 0$, so that the effective probability $\varkappa$ increases with increase of $\alpha$ and decreases with increase of the thickness of F. In the limiting cases, $\varkappa\rightarrow 1$ when $d\rightarrow 0$ and $\varkappa\rightarrow\alpha$ when $d\rightarrow\infty$.

Now we turn to a discussion of the size effect on the magnon heat current in the investigated I/F/I system. From the preceding analysis it follows that in the magnon overheating regime at $\varepsilon\ll1$ both the $Q(T_s)$ dependence and the spectral distribution of phonons emitted by F are determined only by the properties of F and are practically independent of the I characteristics. Thus, in contrast to the ``Joule''  heating regime at $\varepsilon\gg1$, in the case of magnonic overheating regime there is no need to explicitly take into account the mechanisms of the heat removal from F. Accordingly, the results of experiments in the $\varepsilon\ll1$ regime contain information on the magnon-phonon interaction in F. In particular, experiments on the determination of the thermal resistance of an F/I interface should allow for the estimation of the microscopic quantity $l_{ls}(T_s)$. In this regard, we would like to comment on the possibility of experimental realizations of the magnon overheating regime.

The main point, which should be easily grasped from Fig. \ref{f3}, is that the magnon overheating regime at $\varepsilon\ll1$ requires $d \ll l_{ls}$ \emph{in conjunction} with the ballistic propagation of phonons through the F/I interface such that so-called \emph{reversed phonons can be neglected}. From the plot in Fig. \ref{f3} it follows that the magnonic overheating mechanisms can be most easily realized at $\bar\alpha\sim1$. Still, the presence of a small number of reverse phonons can affect experimental results even if the criterion $\varepsilon\ll1$ is formally satisfied. To reduce the number of reverse phonons, in addition to using single-crystal bulk substrates two other experimental possibilities should be mentioned. One is to decrease the width of the investigated F film, to maximally exploit the effect of ``spreading'' the heat into the substrate. Another possibility is to use pulsed heating of F in such a way that the characteristic time for the reversed phonons to enter the film is longer than the duration of the pulse. Still, the pulse duration should be long enough to establish a stationary state in the film. In the case of pulsed heating, the requirements on the width of the film can be greatly relaxed as compared to the case of stationary heating, owing to the lack of a characteristic thermal ``background''  in the pulsed regime.

An experimental criterion to prove that the film is indeed in the magnon overheating regime is the absence of discontinuities in the observed physical quantities when the helium bath temperature goes through the $\lambda$ point. In fact, since the physical characteristics of magnons in F in this regime are no longer dependent on the bare transparency $\alpha$, the inequality $\varepsilon\ll1$ can only be strengthened when the sample is emersed in superfluid helium and, hence, is efficiently cooled. Moreover, if discontinuities exist nevertheless, their magnitude can serve as a measure of the ``distance'' from the magnon overheating regime.

Finally, it is worth noting that evidence for the role of the magnon energy relaxation length in the SSE has been presented in recent work \cite{Pra18prb}, where relaxation processes have been parameterized by the length over which magnon-to-phonon thermalization occurs. At the same time, that model represents a concept, rather than a complete theory of transport, at short thicknesses. Namely, in order to analyze the corresponding contribution to the injected spin current, the magnon heat Kapitza length at the interface is treated in \cite{Pra18prb} as a free parameter and there is no estimate for the interfacial thermal conductance. While in the present work we have been able to microscopically account for the phonon-magnon energy relaxation length at different thicknesses of the F layer and temperatures, a theoretical account for the nonlinear Kapitza resistance of the F/I interface, which exhibits a size effect as well, will be presented in a follow-up work.

\section{Conclusion}

To summarize, we have investigated the heat transfer through interfaces of an insulating ferromagnet sandwiched between two dissimilar insulators. A closed-form expression \eqref{eHeatFluxInt} for the heat current $Q(T_s)$ has been derived at an arbitrary value of the parameter $\varepsilon\sim 2d/\alpha l_{ls}(T_s)$. Depending on the relation between the ferromagnet thickness, $d$, and the mean free path of phonons generated by magnons, $l_{ls}$, two qualitatively different heat-removal mechanisms have been revealed. If $\varepsilon\gg1$, this is the conventional ``Joule''  heating, which has been extensively investigated in a number of works on the thermal resistance between two dissimilar solids. In the case $\varepsilon\ll1$ the magnonic overheating regime becomes possible for the ballistic propagation of phonons through the F/I interfaces. In the magnon overheating regime, the nonlinear in temperature effects are determined only by the properties of F and do not depend on the acoustic transparency of the F/I interface. The predicted magnon overheating regime should be examined in hybrid structures with thin ferromagnetic layers with interfaces exhibiting a good transparency for thermal phonons. In all, the reported results are relevant for the analysis of the spin Seebeck effect hybrid nanostructures and the development of magnon-based spin caloritronic devices.

\begin{acknowledgments}
VAS thanks Dmytro A. Bozhko for a fruitful discussion. Research leading to these results received funding from the European Commission in the framework of the program Marie Sklodowska-Curie Actions --- Research and Innovation Staff Exchange (MSCA-RISE) under Grant Agreement No. 644348 (MagIC).
\end{acknowledgments}

\section*{Appendix}
This appendix provides details on the derivation of the kinetic equation and its solution.

\textbf{Number of phonons}. The change of the number of phonons with a given wave vector $\mathbf{q}$ caused by absorption and emission of a phonon by magnon per unit of time can be presented [Eq. (26.3.1) in \cite{Akh67boo}] as
\begin{equation}
    \label{a16}
    (\dot N_q )_s = L_{ls} \lbrace N,n\rbrace,
\end{equation}
where the right part is the collision integral between phonons and magnons with the distribution functions $N$ and $n$, respectively
\begin{eqnarray}
    \label{a17}
    \begin{array}{lll}
    L_{ls} \lbrace N,n\rbrace =\\
    \frac{2\pi}{\hbar}\sum_{\mathbf{k},\mathbf{k}_1}{|\psi_{sl}(\mathbf{q},\mathbf{k}|\mathbf{k}_1)|}^2
    \lbrace(N_{\mathbf{q}} + 1)(n_{\mathbf{k}} +1) n_{\mathbf{k}_1}  \\
    -N_{\mathbf{q}}n_{\mathbf{k}}(n_{\mathbf{k}_1} + 1)\rbrace\times\delta(\hbar\omega_q+\varepsilon_{\mathbf{k} }
    -\varepsilon_{\mathbf{k}_1})\Delta(\mathbf{q} +\mathbf{k} -\mathbf{k}_1).
    \end{array}
\end{eqnarray}

With the momentum conservation law, Eq. \eqref{a17} reads
\begin{eqnarray}
    \label{a18}
    \begin{array}{lll}
    L_{ls} \lbrace N,n\rbrace = \\
    \frac{2\pi}{\hbar}\sum_{\mathbf{k}}{|\psi_{sl}(\mathbf{q},\mathbf{k}|\mathbf{k} +\mathbf{q})|}^2 \lbrace(N_{\mathbf{q}} + 1)(n_{\mathbf{k}} +1) n_{\mathbf{k} +\mathbf{q}} \\
    -N_{\mathbf{q}}n_{\mathbf{k}}(n_{\mathbf{k} +\mathbf{q}} + 1)\rbrace\times\delta(\hbar\omega_q+\varepsilon_{\mathbf{k} } -\varepsilon_{\mathbf{k} +\mathbf{q}}).
    \end{array}
\end{eqnarray}

Here, ${|\psi(\mathbf{q},\mathbf{k}|\mathbf{k} +\mathbf{q})|}^2$ is expressed by [Eq. (26.1.4) in \cite{Akh67boo}]
\begin{equation}
    \label{19}
    |\psi(\mathbf{q},\mathbf{k}|\mathbf{k} +\mathbf{q})|^2 = \frac{{\Theta_C}^2}{N}(\frac{\hbar}{\rho a^3 \omega_q })a^4 k^2 {(\mathbf{k} +\mathbf{q} )}^2 q^2,
\end{equation}
where $\rho = M/{a^3}$, $M$ is the magnetic ion mass, $a$ is the lattice constant, $\Theta_C$ is the Curie temperature, $N$ is the number of atoms, $\omega_q = s q$ is the phonon frequency with wave vector $\mathbf{q}$, $s$ is the average sound velocity. In the collision integral between phonons and magnons \eqref{a18} $N_{\mathbf{q}}$ and $n_{\mathbf{k}}$ are the Bose-Einstein distributions at the temperature $T$, which in the equilibrium state read $\bar N_{\mathbf{q}} = [e^{\frac{\hbar\omega_q}{T}}-1]^{-1}$ and $\bar n_{\mathbf{k}}=[e^{\frac{\varepsilon_k}{T}}-1]^{-1}$, where $\varepsilon_k=\Theta_C{(ak)}^2$ is the dispersion law of magnons in the long-wave $ka\ll 1$ limit. Obviously, $L_{ls}{\lbrace{\bar N},{\bar n}\rbrace}=0$.

\textbf{Relaxation frequency}. If the number of phonons and magnons is weakly distinguished from the equilibrium case at the temperature $T$, then it is possible to determine the inverse lifetime of a phonon with the frequency $\omega$ with respect to the absorption or emission of the phonon by a magnon by the formulae
\begin{equation}
    \label{A5}
    \upsilon_{ls} = \frac{1}{\tau_{ls}(\omega)}=-\frac{\delta L_{ls}}{\delta N_q (\omega)},
\end{equation}
which yields for the relaxation frequency
\begin{equation}
    \label{A6}
    \begin{array}{lll}
    \upsilon_{ls}(T)= \frac{2\pi}{\hbar}\sum_{\mathbf{k}}(\frac{{\Theta_C}^2}{N})(\frac{\hbar}{M \omega_q })a^4 k^2 q^2(k^2 + 2{\mathbf{k}}{\mathbf{q}} +q^2 )\\[2mm]
\hspace{20mm}\times\lbrace \bar{n}_{\mathbf{k}}-\bar{n}_{\mathbf{k}+\mathbf{q}}\rbrace\times\delta(\hbar\omega_q+\varepsilon_{\mathbf{k}}-\varepsilon_{\mathbf{k}+\mathbf{q}}).
    \end{array}
\end{equation}

In the long-wave limit $ka\ll 1$ the sum is replaced by the integral
\begin{equation}
    \label{A7}
    \sum_{\mathbf{k}}\to\frac{V}{(2\pi)^3}\int{d\mathbf{k}} = \frac{N a^3}{(2\pi)^3}\int dk k^2 dO_k,
\end{equation}
where  $dO = 2\pi\sin\theta d\theta$, where $\theta$ is the polar angle of the vector $\mathbf{k}$ with respect to the vector $\mathbf{q}$. With
$$\delta(\hbar\omega_q+\varepsilon_k-\varepsilon_{k+q})=\frac{1}{\Theta_C(2 a^2 qk)}\delta(f - \cos\theta),$$ where $f=\frac{1}{2ak}(\frac{\Theta_D}{\Theta_C}- qa)$, at low temperatures $T\ll{\Theta_D}^2/\Theta_C$ one obtains
\begin{equation}
    \label{A8}
    \begin{array}{lll}
    \nu_{ls}(T,q)\equiv D(T) J_D (T)= \\[3mm]
    \hspace{10mm}D(T){\int_{y_0}^\infty}dy y (x+y)(\frac{1}{e^y-1}-\frac{1}{e^{x+y}-1}),
    \end{array}
\end{equation}
where $D(T)=\Theta_C(T/\Theta_C)^3/(8\pi MaS)$, $x=\varepsilon_q/T$, $y=\varepsilon_k/T$, $y_0=\Theta_D^2/(4T\Theta_C)$, and $\Theta_D=\hbar s/a$.
The final limit of integration $y_0$ in Eq. \eqref{A8} over the dimensionless magnon energy $y=\varepsilon_k/T$ is due to the fact that the emission of phonons is possible only for magnons with the energy larger than ${\Theta_D}^2/4 \Theta_C$.

In Eq. \eqref{A8}
\begin{equation}
    \label{a13}
    J_D(T_s, x, y_0)= \sum_{p=1}^{\infty} (1-e^{-px})(x\phi _1 + \phi _2)
\end{equation}
where
\begin{equation}
    \label{a14}
    \begin{array}{lll}
    \phi_1(p, y_0 )= e^{-py_0}(y_0/p +1/p^2),\\[2mm]
    \phi _2(p, y_0) = e^{-py_0} ({y_0}^2/p + 2y_0/p^2 + 2/p^3).
    \end{array}
\end{equation}

Because of $J_D(T_s, x, y_0)\sim e^{-2y_0}\ll1$ for $p=2$ we can confine our consideration by $p=1$ in the limit $y_0\gg1$, obtaining
\begin{equation}
    \label{a15}
    \begin{array}{lll}
    J_D(T_s, x, y_0) \approx\\[3mm] \hspace{10mm}(1-e^{-x})e^{-y_0}[x(y_0 +1) +({y_0}^2 + 2y_0 + 2)].
    \end{array}
\end{equation}

\textbf{Collision integral}. The collision integral $L_{ls}\lbrace N, n\rbrace$ in the case, when $N_q$ is arbitrary and $n_k$ is the equilibrium distribution function for magnons with the momentum $\mathbf{k}$ at the temperature $T_s$, reads
\begin{eqnarray}
    \label{a3}
    \begin{array}{lll}
    L_{ls} \lbrace N,n\rbrace = \frac{2\pi}{\hbar}\sum_{\mathbf{k},}\frac{{\Theta_C}^2}{N}(\frac{\hbar}{\rho a^3 \omega_q })a^4 k^2 {(\mathbf{k} +\mathbf{q} )}^2 q^2\\[3mm] \times\delta(\hbar\omega_q+\varepsilon_{\mathbf{k} } -\varepsilon_{\mathbf{k}+\mathbf{q}})\phi (N_q,n_k,n_{k+q})
    \end{array}
\end{eqnarray}
where
\begin{equation}
    \label{a4}
    \phi(N_q,n_k,n_{k+q}) =(N_{\mathbf{q}} + 1)(n_{\mathbf{k}} +1) n_{\mathbf{k} +\mathbf{q}}- N_{\mathbf{q}}n_{\mathbf{k}}(n_{\mathbf{k} +\mathbf{q}} + 1).
\end{equation}
Here, $n_k=[e^{\frac{\varepsilon_k}{T_s}}-1]^{-1}$ and $n_{k+q}=[e^{\frac{\varepsilon_k+\varepsilon_q}{T_s}}-1)^{-1}]$, where $\varepsilon_k = \Theta_C (ak)^2$ and $\varepsilon_q = \hbar\omega_q = \hbar s q$. If we determine  $x\equiv\varepsilon_q/T_s $ and $ y\equiv\varepsilon_k/T_s$, then it can be shown that $n_{\mathbf{k} +\mathbf{q}}(n_{\mathbf{k}} +1)={e^y}/{(e^{y+x}-1)(e^y -1)}$, $ n_{\mathbf{k}}(n_{\mathbf{k} +\mathbf{q}}+1) = {e^{y+x}}/{(e^{y+x}-1)(e^y -1)}$. Then
\begin{eqnarray}
    \label{a5}
    \phi =\frac{e^y [(N_q+ 1)-N_q e^x]}{(e^{y+x} -1)(e^y - 1)}=\frac{e^{y-x}[(N_q +1)- N_q e^x]}{(e^y - e^{-x})(e^y -1)}
\end{eqnarray}
and, finally,
\begin{eqnarray}
    \label{a6}
    \phi =[N_q - \frac{1}{(e^x -1)}][\frac{e^{y+x}}{e^{y+x}-1}-\frac{e^y}{e^y -1}].
\end{eqnarray}
Eq. \eqref{a6} can be written in a more compact form if one presents the two terms in the second bracket as two sums of the geometric series $e^{-(x+y)}$ and $e^{-y}$, namely,
\begin{eqnarray}
    \label{a7}
    \phi =[n(T_s)-N_q]\sum_{p=1}^{\infty} e^{-py}(1-e^{-px}).
\end{eqnarray}
Proceeding, again from summation to integration, at low temperatures $T\ll{\Theta_D}^2/\Theta_C$ we obtain
\begin{equation}
    \label{a9}
    \begin{array}{lll}
    L_{ls} \lbrace N_q,n(T_s)\rbrace =\\[3mm]
    [n(T_s)-N_q]D(T_s)\sum_{p=1}^{\infty} (1-e^{-px}){\int_{y_0}^\infty}y (y+x) e^{-py} dy.
    \end{array}
\end{equation}
Finally, Eq. \eqref{a9} can be written as
\begin{equation}
    \label{a9a}
    L_{ls} \lbrace N_q,n(T_s)\rbrace =[n(T_s)- N_q]\nu_{ls}(T_s, q),
\end{equation}
where
\begin{equation}
    \label{a10}
 \nu_{ls}(T_s, q) = D(T_s)\sum_{p=1}^{\infty} (1-e^{-px}){\int_{y_0}^\infty}y (y+x) e^{-py} dy
\end{equation}
has a physical meaning of the inverse lifetime of a phonon with the frequency  $\omega_q$ with respect to absorption or emission of the phonon by a magnon
\begin{equation}
    \label{eRelaxFreqA}
    \nu_{ls}(T_s, q) = 1/  \tau_{ls}(T_s, q) = - \delta  L_{ls}/\delta N_q
\end{equation}

\textbf{Solution of the kinetic equation}. The general solution of Eq. \eqref{eKinEq} reads \cite{Shk80spj}
\begin{equation}
    \label{eKinEqSola}
    N(z) = C\exp(-z/{l_z}) + n(T_s),
\end{equation}
where $C$ is an arbitrary constant and $l_z = s_z/\nu_{ls}$. We introduce two new functions $N^\gtrless(\mathbf{q},z) = N(z,q,q_z\gtrless 0)$
and denote $| l_z | = l$, where $l$ depends on the angle $\theta$ between the direction of the vector $\mathbf{q}$ and the $z$ axis. We then have for $N^\gtrless$
the relations
\begin{equation}
    \label{eNgglla}
    N^\gtrless(z) = C^\gtrless\exp(\mp z/l) + n(T_s),
\end{equation}
where the coefficients must be determined from the two boundary conditions for $N(z)$ at $z = 0$ and $z = d$, respectively. We consider the case of ballistic propagation of the phonons emitted by F through the F/I boundary, taking into account the finite transparency of the F/I interface within the framework of the acoustic-mismatch theory \cite{Lit59cjp}. We denote the coefficients of the phonon reflection from boundaries 1 and 2 as $\beta_1$ and $\beta_2$. Then, $\beta_i = 1 - \alpha_i, i=1,2$, where $\alpha_i(\theta)$ is the transparency coefficient, and the boundary conditions read
\begin{eqnarray}
    \label{eBonConda}
    N^> (0) & = \alpha_1 n(T_1) + \beta_1 N^< (0), \nonumber\\
    N^< (d) & = \alpha_2 n(T_2) + \beta_2 N^> (d).
\end{eqnarray}

Within the framework of the acoustic-mismatch theory, the interface transparency coefficient $\alpha$ is determined by the phonon incidence angle $\theta_1$ at the interface, the refraction angle $\theta_2$, and the acoustic impedances $z_i$ of the adjacent media via \cite{Lit59cjp}
\begin{equation}
    \label{eTranspCoefa}
    \alpha = \alpha_1 (\theta_1 ) = \alpha_2 (\theta_2 ) = \frac {4(z_2/z_1)(\cos\theta_2/\cos\theta_1)}{[(z_2/z_1) + (\cos\theta_2/\cos\theta_1)]^2},
\end{equation}
where the angles $\theta_1$ and $\theta_2$ at a given boundary are connected via $s_2 \sin\theta_1 = s_1 \sin\theta_2$. Here, subscripts 1 and 2  pertain to one of the F/I interfaces. Combining Eq. \eqref{eNgglla} and \eqref{eBonConda} we obtain the following expression for $C^>$
\begin{equation}
    \label{eC>a}
    C^> = \frac{\alpha_1 n(T_1) + \beta_1 \alpha_2 n(T_2) x + n(T_s)(\alpha_1 -\beta_1 \alpha_2 x)}{1 - \beta_1 \beta_2 x^2}.
\end{equation}
The expression for $C^<$ differs from \eqref{eC>a} by interchanged subscripts 1 and 2, and by an additional factor $x = \exp(-d/l)$. These expressions are used in Eq. \eqref{eHeatFlux}.

\textbf{Heat current}. With the passage from the sum to integration and after the introduction of the magnon ``overheating'' parameter $\gamma\equiv T_s/T_l >1$, the heat current $Q=\sum_{\bf q}(\hbar\omega_{\bf q}) \dot{N}_{\bf q}$ from magnons to phonons acquires the form
\begin{equation}
    \label{eQlonga}
    \begin{array}{lll}
    Q=({N}/{8\pi^3})({\Theta_D^2 \Theta_C}/{2\hbar\Theta_p})({T_s}/{\Theta_C})^3\times\\[3mm]
    \hspace{25mm}[({T_s}/{\Theta_D})^4-({T_l}/{\Theta_D})^4] K(p),
    \end{array}
\end{equation}
\begin{equation}
    \label{a16}
    \begin{array}{lll}
    K(p)=\\[3mm]
    {\int_0}^\infty \displaystyle\frac{ u^3 du}{e^x -1} [J_D (T_s, x=u,y_0)- J_D(T_s, x=u/\gamma,y_0)],
    \end{array}
\end{equation}
The difference in the bracket is given by
\begin{equation}
    \label{a17}
    \begin{array}{lll}
    [...]=
    \sum_{p=1}^\infty u\phi_1[(1-e^{-pu})-(1/\gamma)(1- e^{-pu/\gamma}]+\\[2mm]
    \hspace{20mm
}\phi _2[(1-e^{-pu})-(1-e^{-pu/\gamma})].
    \end{array}
\end{equation}
The dependence of $K(p)$ on the magnon overheating parameter $\gamma$ and the effective inverse temperature $y_0$ is discussed Ref. \cite{Shk18arx}. The corresponding integrals were calculated using Eq. (2.3.13.22) in \cite{Pru02boo}
\begin{equation}
    \label{a16}
    \int_0^\infty \frac{ u^{n-1}(e^{-pu}) du}{e^u -1}= \Gamma(n)[\zeta(n,1+p)].
\end{equation}
For instance,
\begin{equation}
    \label{a17}
    \begin{array}{lll}
    K(p=1)=\phi_1\Gamma(5)[1+\mu[\zeta(5,1+\mu)-\zeta(5)]]+\\[2mm]
    \hspace{30mm}\phi_2\Gamma(4)[1+\mu[\zeta(4,1+\mu)-\zeta(4)]].
    \end{array}
\end{equation}
Here $\mu=1/\gamma=T_l/T_s$, $\Gamma(n)$ is the gamma-function of $n$ and $\zeta (n,1+\mu)$ is the generalized Riman zeta-function of $n$ and $(1+\mu)$, namely
\begin{equation}
    \label{17a}
    \begin{array}{lll}
    \Gamma(n+1)= n!,\hspace{3mm}\zeta(s)=  \sum_{k=1}^{\infty}(k)^{-s},\hspace{3mm}\zeta(s,1)= \zeta(s), \\[2mm]
    \zeta(s,1+p)= \sum_{k=0}^{\infty}[k+(1+p)]^{-s}.
    \end{array}
\end{equation}


\begin{thebibliography}{46}
\bibitem [{1()}]{1}%
  \BibitemOpen
  1,\ \href@noop {} {}\bibinfo {note} {See Fig.~3 in \cite {Sch13prb} where the
  phonon, electron and magnon temperature profiles are shown for an
  insulator/insulating-ferromagnet/normal-metal structure}\BibitemShut
  {NoStop}%
\end{thebibliography}

%

\end{document}